\documentclass[conference]{IEEEtran}
\IEEEoverridecommandlockouts
\usepackage[pdftex]{graphicx}
\usepackage[none]{hyphenat}
\usepackage{amsmath}
\usepackage[]{algorithm2e}
\usepackage{multirow}
\usepackage{booktabs}

\begin{document}
	
\title{Hybrid Deep Learning Model using SPCAGAN Augmentation for Insider Threat Analysis \\}

\author{\IEEEauthorblockN{R G Gayathri, Atul Sajjanhar, Yong Xiang}
	\IEEEauthorblockA{\textit{School of Information Technology} \\
		\textit{Deakin University}\\
		Geelong, VIC 3217, Australia \\
		\{gradhabaigopina, atul.sajjanhar, yong.xiang\}@deakin.edu.au}
}	
\maketitle	
\begin{abstract}
Cyberattacks from within an organization's trusted entities are known as insider threats. Anomaly detection using deep learning requires comprehensive data, but insider threat data is not readily available due to confidentiality concerns of organizations. Therefore, there arises demand to generate synthetic data to explore enhanced approaches for threat analysis. We propose a linear manifold learning-based generative adversarial network, SPCAGAN, that takes input from heterogeneous data sources and adds a novel loss function to train the generator to produce high-quality data that closely resembles the original data distribution. Furthermore, we introduce a deep learning-based hybrid model for insider threat analysis. We provide extensive experiments for data synthesis, anomaly detection, adversarial robustness, and synthetic data quality analysis using benchmark datasets. In this context, empirical comparisons show that GAN-based oversampling is competitive with numerous typical oversampling regimes. For synthetic data generation, our SPCAGAN model overcame the problem of mode collapse and converged faster than previous GAN models. Results demonstrate that our proposed approach has a lower error, is more accurate, and generates substantially superior synthetic insider threat data than previous models.
\end{abstract}

\textbf{Keywords :} class imbalance, insider threat, hybrid deep learning models, generative adversarial networks, adversarial training, adversarial robustness

\section{Introduction}
Organizations are always at high risk from various kinds of cyberattacks. The attacks originated within the organization, referred to as insider attacks, where attackers are in close association with the workplace directly or indirectly and physically or logically are of serious concern. Insider attacks have been on the rise recently, prompting experts to consider them more seriously. According to a survey conducted by Gurucul \cite{gurukul} in 2020, insider attacks have increased by 67\% in the last two years. Lack of real-world data and data imbalance leaves insider threat analysis an understudied research area. Many machine learning methods with cutting-edge performance rely heavily on data augmentation \cite{paschali2019data} to mitigate data shortage. 

Common augmentation strategies operate by randomly selecting samples from a space of transformations. However, even if an augmentation strategy improves a dataset in a specific application does not indicate it can be successfully applied to other datasets and applications \cite{cubuk2019autoaugment}. Unfortunately, due to the curse of dimensionality, such sampling approaches are limited in expressiveness because they cannot scale to complex transformations that depend on numerous parameters. Adversarial examples \cite{goodfellow2014explaining} can be thought of as a different strategy for data augmentation. By being trained on the challenging input modifications, the resulting models should manage other, apparently simpler, modifications. The benefit of adversarial augmentation is that it replaces sampling with a single, calculated perturbation that provides the maximum loss.

The majority of insider detection techniques are anomaly-based, which means they look for unusual activities, including those that have never happened before, since they consider any unusual situation a potential attack. Classification using neural networks is a popular approach used in anomaly detection \cite{chandola2009anomaly}. In contrast to traditional neural networks, a Bayesian Neural Network does not overfit on small datasets \cite{popkes2019interpretable}. Nowadays, the deep neural networks (DNN) trained with dropout is interpreted as a Bayesian technique. Consequently, Bayesian Neural Networks (BNNs) have recently gained much attention.



In this paper, we combine the potential of Generative Adversarial Networks(GANs) \cite{goodfellow2014generative} to perform data augmentation and hybrid model using Bayesian Neural Networks(BNNs) for improved anomaly detection. We propose a modified Auxiliary Classifier Generative Adversarial Network (ACGAN) \cite{odena2017conditional}, referred to as SPCAGAN, to generate high-quality data and to improve the performance of insider threat analysis in end-to-end models. We adopt a linear intrinsic dimension estimation method to study the data distribution in feature space. When properly trained, the GAN generator can efficiently generate perturbations for each instance, thus speeding up adversarial training as a defense.

To the best of our knowledge, this paper is the first attempt to use the GAN networks with linear intrinsic dimension estimation methods in insider threat analysis to achieve improved performance. We propose a manifold-learning-based adversarial training approach for insider threat analysis using the generative models. Our goal is two-fold : (i) reduce the adverse effects of the class imbalance by GAN-based data augmentation (ii) build effective anomaly detection models resistant to adversarial attacks. Exhaustive experiments are performed based on a benchmark dataset under various settings to demonstrate the effectiveness of GAN in insider detection in comparison with the existing works.

The contributions of the paper are : 
\begin{itemize}
	
	\item We developed a linear manifold learning-based regularization for adversarial training that helps in improved insider threat analysis.
	
	\item We propose an anomaly detection model based on probabilistic deep learning. This model shows the feasibility of implementing a hybrid model using Bayesian methods and other deep learning models.
	
	
	\item We present a systematic evaluation that gives a detailed analysis of how the proposed approach improved the performance in terms of popular measures compared to the existing works on benchmark datasets. Experiments show that our approach can effectively achieve more accurate anomaly detection than state-of-the-art models.
\end{itemize}

The rest of this paper is organized as follows. Section \ref{sec_Lit_Review} summarizes the existing insider threat approaches using machine learning (ML) and deep learning (DL) methods and synthetic data generation methods. Section \ref{sec_methodology} presents a technical overview of the proposed approach. Section \ref{sec_exp_setup} briefs about the data description being used for validation and the algorithms used, followed by an evaluation of the performance of our method in terms of various metrics. Section \ref{sec_conclusion} concludes the paper and identifies potential future research directions.

\section{Related Work}\label{sec_Lit_Review}
This section reviews relevant literature on the state-of-the-art insider threat approaches, the potential of GAN-based data augmentation, and the increased relevance of hybrid learning models to position our research with existing research.

\subsection{Insider Threat Analysis} \label{sec_rel_work_insider}
Insiders are individuals such as employees, contractors, or business partners of an organization who are trustworthy sources, have legitimate access rights, and execute suspicious behaviors within the organization, either purposefully or accidentally. Unintentional activities sometimes occur due to negligence or a lack of understanding of security policies. Although insider threat analysis has been studied for many years, the emergence of new approaches is limited due to a lack of real-world data and difficulty finding an effective solution.

Nowadays insider threat analysis attracts wider attention due to the increase in frequency of attacks and the advent of promising data analysis techniques. Existing works helped to understand the exhaustive range of methods applied from various perspectives. To list a few, it includes statistical metric based solutions \cite{meng2020detecting}, machine learning \cite{le2020analyzing}, 
image-based analysis \cite{gayathri2020image}, 
blockchain-based analysis \cite{meng2019enhancing}, 
natural language processing (NLP) approaches\cite{liu2019insider} 
and dealing with the insider attack in several critical application areas like health care \cite{meng2018towards} and Internet of Things \cite{khan2019malicious}. There is an upward trend towards using machine learning and deep learning based solutions for insider threat analysis. 

In the survey \cite{homoliak2019insight}, Homoliak et al. presented a classification of insider threat research publications and promising study directions. The authors identify the use of GAN for adversarial classification for enhanced threat detection and data augmentation while recommending possible research directions.

Existing works use various data generation and augmentation strategies in insider threat detection. The method followed in \cite{soh2019employee} based on the email communications create a new set of data, Enron\textsuperscript{+}, followed by anomaly detection using Isolation Forest algorithm. The scenario-based classification in the work \cite{chattopadhyay2018scenario} considered the problem as three different binary classification problems and used Random Over Sampling (ROS) for data augmentation under various sampling ratios. The method gave good results for Scenario 2 on Random Forest whereas it could not perform satisfactorily on Scenarios 1 and 3. Spread subsample, a method that selects the random samples to fit in the memory by balancing the skewed class distributions is mentioned in the paper \cite{resample_weka}. An ensemble strategy combined with data adjusted XGBoost model for insider threat detection proposed in \cite{smote_xgb} gave promising results. All these works performed supervised binary classification for insider detection. In the paper \cite{trustcom}, the authors considered class labels also in the data generation process which helped in generating more meaningful data. 



\subsection{GAN-based Data Augmentation}
Unlike other synthetic data generation methods, GANs can create realistic data samples that significantly reduce data imbalance and ultimately overfitting. Since the network uses the original data and noise to create synthetic data, it can be considered a transformation approach where generated data points mimic the real data distribution. GANs are very popular with image processing applications. 
Very recently, GANs have been applied in cybersecurity \cite{piplai2020nattack}, \cite{lee2021gan} which is a source of heterogeneous data in structured and unstructured form.

In the paper \cite{piplai2020nattack}, the authors used generative adversarial networks in intrusion detection. Lee and Park \cite{lee2021gan} proposed a method for detecting intrusion using GAN and random forest algorithm. The model was built to solve overfitting. The experiments show that the model achieved its goal and performed well in imbalanced datasets. The authors have proposed a conditional GAN \cite{zhang2020network} for insider threat detection, which proved to be effective for multiple attack detection with many minority classes.

Most of the existing GAN research concentrates on unstructured, continuous data such as images. However, real-world datasets used for classification purposes are mostly tabular and include numerical and categorical attributes. GAN-based augmentation methods are gaining popularity 
\cite{tran2021data},
\cite{arantes2020csc}, 
\cite{zhu2018emotion}. GAN-based approaches for modeling tabular data have recently been developed, whereas GAN-based approaches for numeric data are still in the budding stage.

Tabular data synthesis comprises a variety of methodologies based on the types of data. It models a joint probability distribution of columns in a table to create a realistic synthetic table. However, due to the limitations of several of these models, such as distributions and processing issues, high-fidelity data synthesis has proven to be challenging. Several GAN-based data generation algorithms for generating synthetic tabular data, notably for healthcare records, have been developed in recent years. 
These methods can be applied on the numeric feature vectors to generate more samples in various application domains. In the paper \cite{saxena2021generativesurvey}, the authors mention the various research directions possible in GAN-based analysis; the use of regularization and the need for modified architectures are a few among them.

\subsection{Deep Learning Models}
Recently, deep learning methods gained attention in insider threat analysis. Yuan and Wu \cite{yuan2021deep} performed an exhaustive review on the challenges and the potential research directions for deep learning in insider threat analysis and pointed out extreme class imbalance as one of the key challenges. Deep neural networks (DNNs) are a popular forecasting tool due to their versatility and ability to represent non-linearities despite the fact they tend to overfit and hence are not recommended for small data sets \cite{salman2019overfitting}. They also don't measure the degree of uncertainty in the projections. 

Bayesian Neural Networks (BNNs) can overcome the above mentioned problems using prior distributions on the parameters of a DNN model and expressing uncertainty about the predictions in the form of a distribution \cite{popkes2019interpretable}. Rather than representing each estimated parameter as a single point, they express it as a distribution. The field of Bayesian Deep Learning, which uses the Bayes technique for neural network models, is fast-growing \cite{bayesiansurvey}, \cite{blundell2015weight}. 

Ensemble techniques \cite{ho2002multiple}, \cite{hansen1990neural} combine multiple machine learning algorithms to provide better predictive performance than a single machine learning classifier. The primary idea behind an ensemble approach is to mix various machine learning algorithms to make use of the strengths of each one to create a more robust classifier. Ensemble techniques are beneficial when the problem can be divided into sub-problems, and each sub-problem is given to one of the ensemble's modules. Each module can comprise one or more machine learning algorithms, depending on the ensemble approach's structure. The most challenging part of adopting ensemble techniques is deciding which collection of classifiers to use to make up the ensemble model and which decision function to use to integrate the outputs of those algorithms.

%


Inspired by the success of GAN as an oversampling method for data augmentation, this article proposes a GAN-based data augmentation insider detection model referred to as SPCAGAN. The model aims to reduce class imbalance issues from the data level. First, SPCAGAN is used to generate specified minority classes followed by merging the generated data into the original training data to create a new training dataset. The proposed model increases the sample diversity and alleviates the class distribution imbalance of the dataset. Second, the new training dataset is used to perform insider threat detection using hybrid Bayesian neural network classification algorithms. We explain the proposed approach in detail in the forthcoming section.

\section{Proposed Method}\label{sec_methodology}

\begin{figure}[!hbtp]
	\centering
	\includegraphics[scale=0.35]{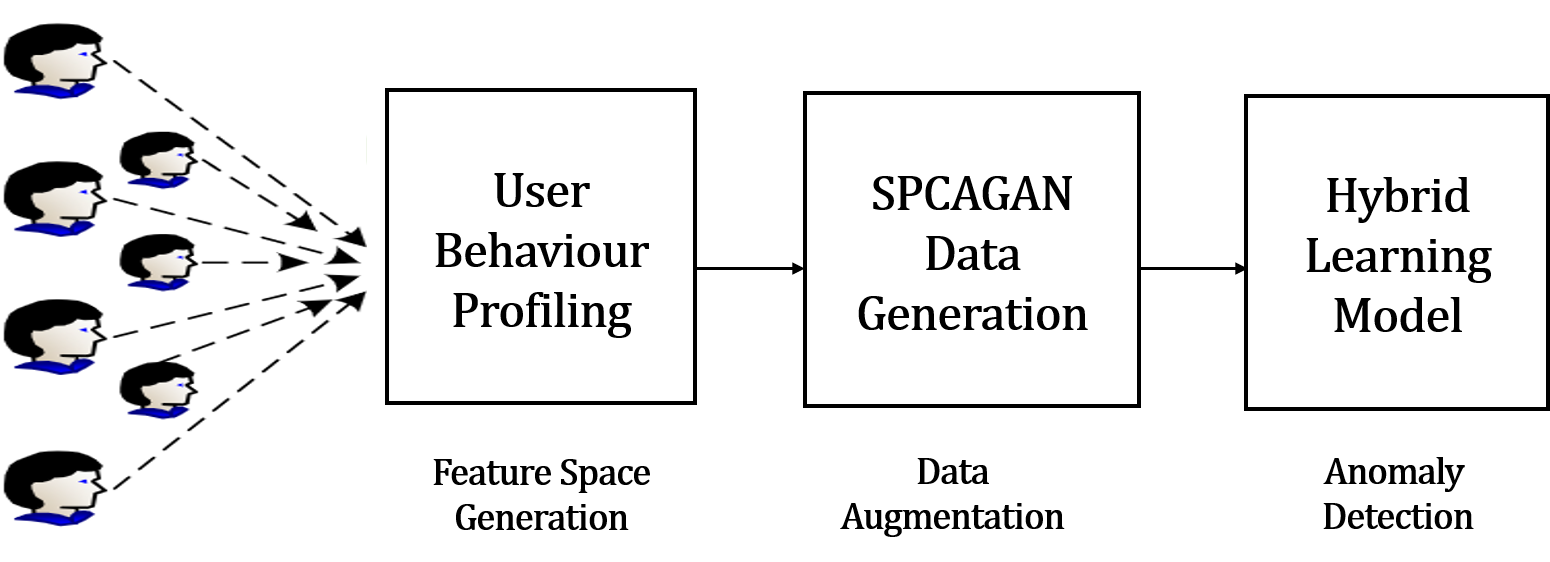}
	\caption{Multi-stage Process flow in the Proposed Approach}
	\label{fig:processflow}
\end{figure}

In this section, we provide a detailed explanation of the proposed method for data augmentation, and insider threat detection. We follow a multi-stage workflow to perform the insider threat analysis. The overall process is split into three sequential stages: (i) User behavior representation (ii) Data Augmentation (iii) Anomaly detection. The first step performs the pre-processing and feature space generation, followed by data augmentation in the second step. Finally, anomaly detection is carried out. Each step is explained in detail in the coming sections. Fig. \ref{fig:processflow} outlines the workflow of the method being used.

Each step involved in the process is explained in detail in the following sections. 

\subsection{Behaviour Extraction}\label{sec_ftr_engg}
Insiders are entities that perform harmful activities from within the organization using various resources to which they have access. Insider attacks can be intentional or unintentional, but the consequences can be severe. Insider activities typically leave a minimal trace in the audit data trails. A variety of behavioral characteristics can identify malicious insiders. Copying significant volumes of data to external/USB drives, transferring confidential material to own personal account via emails, abnormal or untimely access patterns on official resources, and using unlawful, abusive, or unethical language in emails or papers are only a few of them.

This section deals with the feature space generation for the insider threat analysis. Each user is identified as an insider based on the entire activity log instead of those related to the particular scenario. User behavior profiling is one of the significant steps in insider threat detection, and this work uses context-based user profiling, where all the features contribute to user behavior. Scenario refers to the sequence of activities that result in a malicious event. The distinct feature sets identified for each scenario are merged to form a single feature set representing context-based features. This makes the feature space easy to scale in cases of new scenario identification.

Growth in technology led to replacing traditional workplace systems with emails, electronic files, websites, and information sharing mediums like external devices, demands for changing conventional security methods to track the employees. Recently, artificial intelligence-based solutions are more popular due to their capability to study historical and current data patterns and develop valuable and usable learning to deter and prevent these kinds of activities.


The various user activities used to represent user behavior are derived from log files. The user activity logs obtained from the logon-logoff details, file access history, external device usage patterns, email communications, and the web browsing history files are pre-processed. We try to incorporate features from each resource mentioned above. Features can be generated based on the scenarios defined in the CERT standards \cite{cmucertscenarios}. 

The features extracted from the resource files are used for threat intelligence techniques. The existing works use numerous feature sets engineered from various perspectives, which are being used as in the machine and deep learning models. As the number and values being used in the features significantly impact the learning models, we performed a step-by-step analysis of the number of features for the user behavior profile. Using the features from the paper \cite{chattopadhyay2018scenario}, we added features from different resources followed by feature selection to derive the effective feature list.

The features newly introduced in this work include the sentiment analysis of the text available in the resources like email and web access files helps in understanding the user sentiment in the communication. Moreover, we included the psychometric details of the users, which hints about the user resource usage in various contexts. We extended the feature list by adding sentiment analysis, holiday details, in-depth feature identification from the email, and web browsing history. Later, we eliminated the less meaningful and dependent features using the feature selection methods.

In the review paper, Guyon and Elisseeff \cite{guyon2003introduction} presents the need for feature selection that acts as a catalyst in the model building in many ways like reducing the training time, helping in selecting a simpler model, reducing the dimension, reducing the chance of overfitting thereby enhancing the model generalization. Non-multicollinearity of predictor variables is assumed in the most linear analysis, which means that pairs of predictor features cannot be correlated. The removal of correlated features is a fundamental feature selection strategy. Multi-correlated features also indicate duplicate features in the data; therefore, removing them is an appropriate starting point when minimizing potential dimensions. 

We can address the challenges of feature redundancy and predictor collinearity by eliminating correlated features, thereby only preserving one of the groups of observed associated characteristics. Finally, the feature list consists of 47 features from all the available files. The feature engineering detailed in this section is not restricted to any particular dataset but can be used for any data with minimal adaptation depending on the availability of the log files. Table \ref{tab:feature_details} gives the details of features extracted from the resource access logs.

\begin{table}[!hbtp]
	\centering
	\caption{Features extracted from the heterogeneous resource files}
	\label{tab:feature_details}
	\begin{tabular}{@{}lccl@{}}
		\toprule
		\textbf{Resource}                                                       & \textbf{\begin{tabular}[c]{@{}c@{}}Feature \\ Count\end{tabular}} & \textbf{\begin{tabular}[c]{@{}c@{}}Data \\ Type\end{tabular}} & \multicolumn{1}{c}{\textbf{\begin{tabular}[c]{@{}c@{}}Method \\ Used\end{tabular}}}                                 \\ \midrule
		\textbf{Logon/Logoff}                                                   & 10                                                                & Numeric            & Statistical                                                              \\
		\textbf{\begin{tabular}[c]{@{}l@{}}Email \\ Communication\end{tabular}} & 12                                                                & Numeric            & \begin{tabular}[c]{@{}l@{}}Statistical\\ \begin{tabular}[c]{@{}l@{}}Sentiment\\ Analysis\end{tabular}\end{tabular} \\
		\textbf{Web access pattern}                                             & 8                                                                 & Numeric            & \begin{tabular}[c]{@{}l@{}}Statistical\\ \begin{tabular}[c]{@{}l@{}}Sentiment\\ Analysis\end{tabular}\end{tabular} \\
		\textbf{File Access}                                                    & 5                                                                 & Numeric            & Statistical                                                              \\
		\textbf{External Device Access}                                         & 5                                                                 & Numeric            & Statistical                                                              \\
		\textbf{Psychometric}                                                   & 5                                                                 & Numeric            & NA                                                                       \\ \bottomrule
	\end{tabular}
\end{table}

The feature set comprehensively describes each user's daily resource usage pattern, which results in a highly imbalanced class distribution due to the rarity of insider attacks. The data samples are labeled based on the scenario from the user's activity pattern. In this work, our main aim is to use a generative model to balance the data distribution and build a hybrid learning model for anomaly detection. The following section explains how generative adversarial networks are utilized to generate meaningful data samples that help to reduce the skewed class distribution.

\subsection{Adversarial Linear Manifold Learning}\label{sec_gen_syn_data}
Data augmentation means expanding the quantity and variance of a dataset used to train a machine learning model to improve generalizability and gain a better understanding of the underlying distribution of the training data. The space that represents the training data distribution can be viewed as the manifold of a class learned by a classifier. Here, we introduce the generation of adversarial data samples for insider threat analysis.

When presented with massive, multidimensional data sets, it is frequently beneficial to identify or impose some structure on the data. The smallest number of parameters M required to explain the data is one such structure. This value M is the data set's intrinsic dimensionality (ID), or more precisely, the data generation process. According to the geometric interpretation, the complete data set is contained within a topological curve of M or fewer dimensions. It is required first to compute the intrinsic dimension of the data set to build a mapping connection from the original data to the reduced feature space. This mapping can be linear/nonlinear and explicit/implicit based on the domain where it is used.

Approaches for estimating IDs that have been created in the past can be categorized into two broad categories: local methods and global methods. Local methods calculate the topological dimension of data using data from sample neighborhoods. Global methods, such as projection, multidimensional scaling, fractal-based analysis, and multiscale analysis, use the complete data set, considering that it is contained within a single manifold of defined dimensions.

The projection method involves projecting data into a low-dimensional space and then confirming the low-dimensional representation of data to obtain the ID. PCA is a standard projection approach that counts the number of significant eigenvalues to find the ID. The intrinsic dimensionality of a data set may be critical in determining the characteristics of classifiers applied to it and, as a result, in selecting the best classifier. These techniques are classified as linear or nonlinear. Principal component analysis (PCA) is a commonly used linear technique discussed below.

S\textsubscript{PCA}, devised by Krzanowski \cite{krzanowski1979between}, is a method for assessing the similarity of two datasets using a PCA similarity factor. The PCA Similarity Factor (S\textsubscript{PCA}) compares two matrices with the same number of columns but not the same number of rows. S\textsubscript{PCA} first computes the principal components of each matrix and then uses heuristics to choose the first $ k $ principal components. For example, the top $ k $ principal components are chosen because their variances account for 95\% of the overall variation. The SPCA algorithm then calculates the similarity of the first k principal components. The PCA Similarity Factor, S\textsubscript{PCA}, between two matrices, A and B, is determined using the angles between principal components and is defined as follows.


\begin{equation}
	S\textsubscript{PCA}(A,B) = trace(LM^{T}ML^{T}) =  \sum_{i=1}^{k} \sum_{j=1}^{k} \cos ^2 \Theta _{ij}
\end{equation}

where L and M are the loadings matrices of the first \textit{k} principal components of A and B, respectively.  
The coefficients of the linear combination of the original variables from which the principal components (PCs) are derived are known as PCA loadings. This value can be used to compare the overall similarity of two spaces and is clearly between $ k $ and $ 0 $, where k is the number of PCs. The range of S\textsubscript{PCA} is between $ 0 $ and $ k $. The advantage of using this method is that the data size is not so important; it is based on the principal components and the distance between those vectors in the subspace. Hence, can be more reliable than correlation-based ones. We exploit this metric for regularizing the GAN training.


Adversarial training \cite{goodfellow2014explaining} is a helpful approach for making neural network models more robust. This technique's core concept is to apply well-known attack methods to generate adversarial samples during the model training stage. Adding the adversarial samples to a training set and retraining until a new model that is resistant to perturbations is built. The robustness and accuracy of the newly created model can be improved using this method. However, the synthesis of different classes of adversarial samples, which increases the amount of data in the training set, can improve the model's generalization and accuracy.


A generative adversarial network (GAN) is composed of two neural network modules, a generator and a discriminator. The objective function of the real GAN commonly referred as vanillaGAN is:

\begin{align*}&\hspace {-.5pc} \min \limits _{G} \max \limits _{D} V(G,D)=\min \limits _{G} \max \limits _{D} E_{{ \boldsymbol { x}}\sim p_{r}} [\log D({ \boldsymbol { x}})] \\& \qquad\quad \qquad \qquad \qquad \qquad\displaystyle { +E_{{ \boldsymbol { z}}\sim p_{z}} [\log (1-D(G({ \boldsymbol { z}})))] } \tag{1}\end{align*}

In the above equation, z represents random noise, p$ z$ represents the distribution of noise samples z, p$ r$ represents the distribution of real data x, p$ g$ represents the distribution of attack samples generated by G, G(z) represents the pseudo data generated by generator G, and E(.) represents the expected value. The two networks compete against each other and are iteratively optimized, with D maximizing the accuracy of differentiating data sources and G generating more realistic fake samples to fool the discriminator D.

Many GAN variants like CGAN \cite{mirza2014conditionalgan}, WGAN-GP \cite{gulrajani2017wgangp}, ACGAN \cite{odena2017conditional} are used in many applications. The vanillaGAN framework, on the other hand, lacks control over the generated data because it is an unconditioned generative model. In other words, the labels for the created images cannot be predicted. Mehdi et al. \cite{mirza2014conditionalgan} presented the conditional generative adversarial network (CGAN) as a way to incorporate label information into the training process and train a generator that produces data from the given class. To produce data under the desired conditions, CGAN was proposed. 

The CGAN generator generates data with the specified condition by taking a pair of latent vectors and condition vectors as input. The generator estimates the distribution of P$(X|y)$ and the discriminator learns to estimate D(X, y) = P(fake\textbar X, y). Following that, ACGAN was proposed as an enhanced version of CGAN. ACGANs are also capable of controlling the output with additional input, but its architecture is more complicated. 

The stability in the training process and its ability to generate high quality synthetic data motivated us to chose ACGAN for inside threat data augmentation. However, for insider threat assessments, the direct application of ACGAN for data augmentation does not give adequate results as seen in the results explained in Section \ref{sec_results}. Different regularization procedures can be considered. We describe a variation that strongly integrates a regularization factor into the optimization technique as part of our attempt to synthesize the features of real-world data. 

\begin{figure}[tbph!]
	\centering
	\includegraphics[scale=0.3]{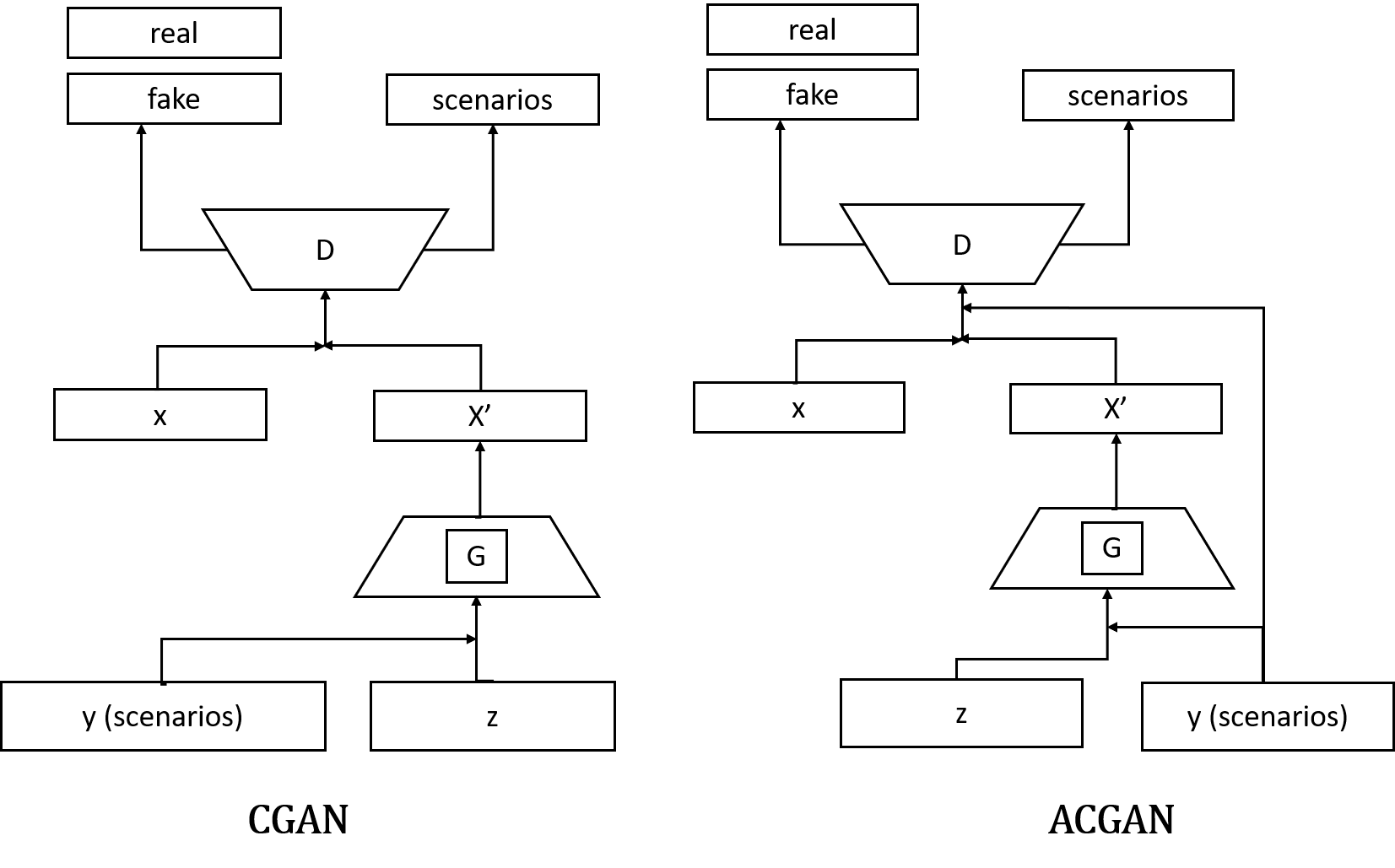}
	\caption[]{Architecture of CGAN and ACGAN}
	\label{fig:cganvsacgan}
\end{figure}

As with any GAN network, the ACGAN has a generator and a discriminator. Like CGAN, the ACGAN's generator model takes a point in the latent space and a class label as input to condition the synthetic data generation process. Unlike the CGAN discriminator, which accepts both data and class labels as input, the ACGAN discriminator accepts real and synthetic data samples. The discriminator model must then predict whether the given image is real or fake, just the same as CGAN does, as well as the data sample's class label. Fig. \ref{fig:cganvsacgan} shows the difference in the CGAN and ACGAN architecture.

The discriminator and auxiliary classifier are independent networks in theory, but in fact, a single NN performs both tasks and produces two outputs. The first output is a single probability obtained from the sigmoid activation function, which denotes the input image's realness and is optimized using binary cross-entropy, just like a standard GAN discriminator model. Like any other multi-class classification neural network model, the second output is a probability of the image belonging to each class obtained using the softmax activation function, optimized using categorical cross-entropy.


In ACGAN, the discriminator gives not only a probability distribution over the real source, but also a probability distribution over the relation labels. The loss function of the ACGAN is divided into two parts : (i) The log likelihood of the correct source — Source-loss (L\textsubscript{S}) as in Equation \ref{eq_ls} (ii) The log likelihood of the correct class — Class-loss (L\textsubscript{C}) as in Equation \ref{eq_lc}. 

\begin{equation}\label{eq_ls}
	L_{S} = E[log P(S = real|X_{real})] + E[log P(S = fake|X_{fake})] 
\end{equation}

\begin{equation}\label{eq_lc}
	L_{C} = E[log P(C = c|X_{real})] + E[log P(C = c|X_{fake})] 
\end{equation}

The generator and discriminator compete over this loss function. The class-loss is maximized by both the generator and the discriminator. The source-loss, on the other hand, is a min-max problem. The generator tries to fool the discriminator by minimizing the source-loss. On the other hand, the discriminator aims at maximizing source-loss while preventing the generator from acquiring an advantage.

Given the relevance of insider threat analysis, we develop a generative model for data augmentation based on ACGAN in which the insider threat scenario labels \textit{y}, as well as the noise sample \textit{z}, are incorporated into the minmax optimization process. This study proposes SPCAGAN, a variation of the auxiliary classifier GAN by incorporating the linear manifold learning to maximize the similarity of real and fake data. 

In this work, we use the manifold learning with respect to real and fake data. Manifold learning is about a class of algorithms to describe that low-dimensional, smooth structure of your high-dimensional data. Depending on the number of dimensions involved, a linear manifold can be considered a line, a plane, or a hyperplane. Data is usually represented in a high-dimensional space, and it is generally expected that a lower-dimensional linear manifold can effectively summarize the relationship between the variables. Most statistical theories and applications dealing with dimensionality reduction focus on linear dimensionality reduction and, by extension, linear manifold learning.


A linear manifold is a simple linear approximation to a more complex form of nonlinear manifold that can probably fit the data better. In both cases, the linear manifold's intrinsic dimensionality is assumed to be significantly lower than the dimensionality of the data. The suggested method for achieving linear dimensionality reduction is to construct a smaller set of linear transformations of the input variables. Linear transformations are projection methods; therefore, the aim is to construct a sequence of low-dimensional projections of the input data with optimal qualities.

\begin{figure}[tbph!]
	\centering
	\includegraphics[scale=0.35]{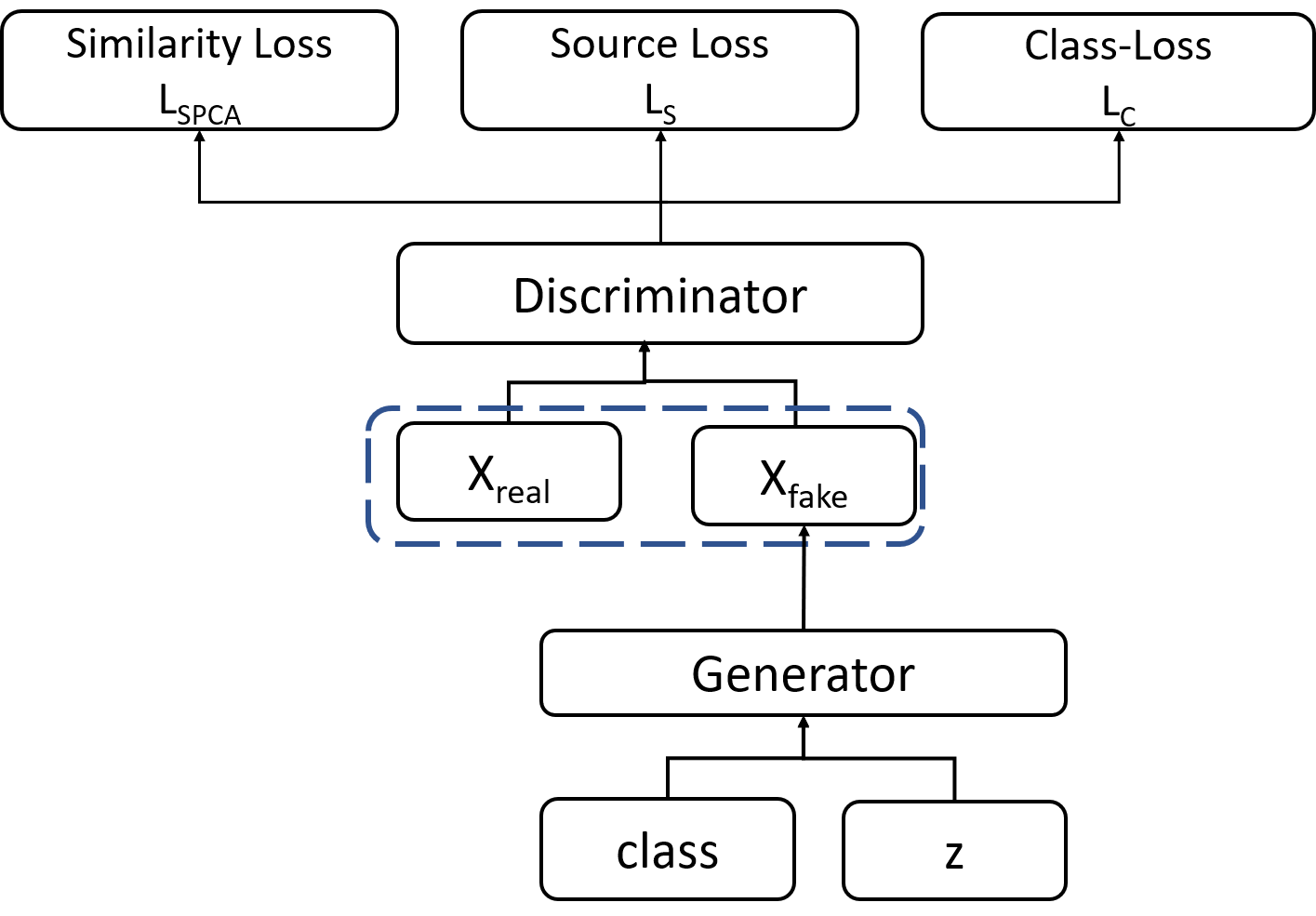}
	\caption{SPCAGAN architecture}
	\label{fig:spcagan}
\end{figure}

The challenge relies first in creating a good representation, which usually converts a high dimensional space, e.g. images, into a lower dimensional representation. Once we have such representation, we would need a technique to map new data samples into the representation, which we will refer as mapping to the latent space. Finally, a metric is applied in the latent space to determine the closeness of unseen data to the learned manifold insider scenarios. The proposed SPCAGAN architecture is given in the Fig. \ref{fig:spcagan}.

As seen in the Fig. \ref{fig:spcagan}, the generator loss is updated with the PCA similarity factor of the real and fake data samples to assure the distribution matching. We propose SPCAGAN which uses a loss function for regularization based on the PCA similarity factor \cite{krzanowski1979between} between real \textit{x} and fake \textit{$\widehat{x}$} as in Equation \ref{eq_spca}. 

\begin{equation} \label{eq_spca}
	L_{SPCA} = E_{x \sim p^{_{data}},\widehat{x} \sim  p^{_{gen(x))}}}[SPCA(x,\widehat{x})]
\end{equation}

This loss is minimized to assure the model regularization for data distribution matching. We estimate SPCA using a batch-based approach: for one batch during training, we use the GAN generated samples as the Z for that batch to compute the SPCA score for the original sample in the batch. In other words, the SPCA(real, fake) is defined efficiently across a set of original and synthetic samples. The algorithm for SPCAGAN training is given in Algorithm \ref{algo:timeseries}.

\RestyleAlgo{ruled} 
\SetKwInput{KwInput}{Input}
\SetKwInput{KwOutput}{Output}
\begin{algorithm}
	\KwInput{Training samples from under represented data, maximum epochs, learning rates}
	\KwOutput{Trained Generator G}
	Initialize generator G and discriminator D \\
	i = 0
	\While{$i < max\_epochs$}{
		\ForEach{minibatch}{
			/*Perform Adversarial Training*/ \\
			Train the discriminator D with minibatch \\
			learning rate = $\lambda_{D}$, loss = $L_{S} + L_{C}$\\
			Train the generator G with mini-batch \\
			learning rate = $\lambda_{G}$, loss= $L_{S} + L_{C} + L_{SPCA}$  \\		
		}
		i = i + 1
	}
	return trained generator G
	\caption{SPCAGAN Training}
	\label{algo:timeseries}
\end{algorithm}

We intend to modify the training process in such a way that the GAN training takes into consideration the manifolds in the data so that the similarity of the data an be maximized. The discriminator is modified in such a way that the auxiliary classifier still works on the classification whereas the data similarity is computed using SPCA. Unlike previous augmentation approaches that use random transformations, SPCAGAN augmentation assures that the network's training area is not restricted to the direct vicinity of a training sample. 

\textit{Network Architecture:} The number of layers used in generator and discriminator are fixed to five. The number of nodes in generator at each hidden layer is {32, 64, 64, 128, 512, 1024}, the discriminator uses the same number of hidden nodes but in a descending order of size. The two neural networks were designed using dense layers. In generator, each layer's activation function was Leaky ReLU, with the exception of the output layer, which used linear activation. The discriminator also uses Leaky ReLU at each dense layer. The output contains two layers : a sigmoid and a sparse categorical cross entropy. The proposed loss function is used in the SPCAGAN model while updating the generator. Once the training is complete, the generator model can be used to create any number of synthetic data for any required class. The next section explains how to use the SPCAGAN generated data for anomaly detection.

\subsection{Anomaly Detection using hybrid Bayesian learning model}
This section discusses our proposed hybrid model, which integrates deterministic and probabilistic methods to perform anomaly detection. Insider threat analysis using anomaly detection using ML/DL algorithms is a well-established method. Existing methods mainly focused on supervised learning with two classes: malicious and non-malicious. The scarcity of insider threat datasets and the diversity of malicious actions are two main reasons for binary classification-based analysis. Deep learning algorithms require a large number of samples in their training dataset to learn the representations and build a robust classifier, but tree-based models are known to work well on minimal data. Despite techniques such as XGBoost that perform well on unbalanced data, the need for more complicated learning models such as artificial neural networks continues to grow. We used a combination of ensemble methods and artificial neural networks, which have recently gained prominence.

Unlike existing methods, the proposed method considers the pre-defined scenario-based malicious activities and gives a deeper insight into the employee activities. Anomaly detection considers the different scenarios under malicious class and normal activities as non-malicious instances and performs the multi-class classification. The classifier tries to discriminate the anomalous activities from the other samples. 

Probability distributions are used to express all uncertain elements in a model, including structural, parametric, out-of-distribution, and noise-related variables and their relationships to data. Deep neural networks with probabilistic layers that can express and process uncertainty are known as probabilistic neural networks. Bayesian Neural Networks (BNNs) are deep neural networks that control overfitting using posterior inference. The key concern with Bayesian neural networks is that their architecture makes computing for uncertainty in many subsequent layers unnecessary and expensive. 

The recommendation in \cite{chang2021bayesian} is to utilize a few probabilistic layers at the network's conclusion. The learning technique can be significantly simplified by training only a few probabilistic layers, and it solves numerous uncertainty-related issues like design and training while still producing valuable results from a Bayesian neural network perspective. It can be regarded as learning a hybrid BNN that includes parts of a standard neural network and shallow BNN. We designed a hybrid model that merges multiple deep learning models as feature extractors, followed by Bayesian network classification.

Gal and Ghahramani \cite{gal2016MCdropout} recently proposed employing Dropout \cite{srivastava2014MCDropTest} at test time to evaluate predictive uncertainty using Monte Carlo dropout (MC-dropout). 
The ease of implementation with MC-dropout has led to its widespread use in practice. Dropout can also be understood as an ensemble model configuration that averages predictions among many NNs. Any reasonable approximation to the true Bayesian posterior distribution must depend on the training data. Hence, the ensemble interpretation sounds more realistic, especially when the training data does not affect dropout rates. This view motivates the usage of ensembles as an alternative method for evaluating predictive uncertainty. 

A feasible solution is to employ hybrid Bayesian neural networks, which use a few probabilistic layers judiciously positioned in the networks. Our research uses a hybrid technique Bayesian Neural Networks, which takes advantage of the strengths of Bayesian inference in Artificial Neural Networks. Different classifiers are used in hybrid approaches, and their predictions are combined to train a meta-learning model. The hybrid system helps to improve the performance of a specific system.

\begin{figure}[tbph!]
	\centering
	\includegraphics[scale=0.35]{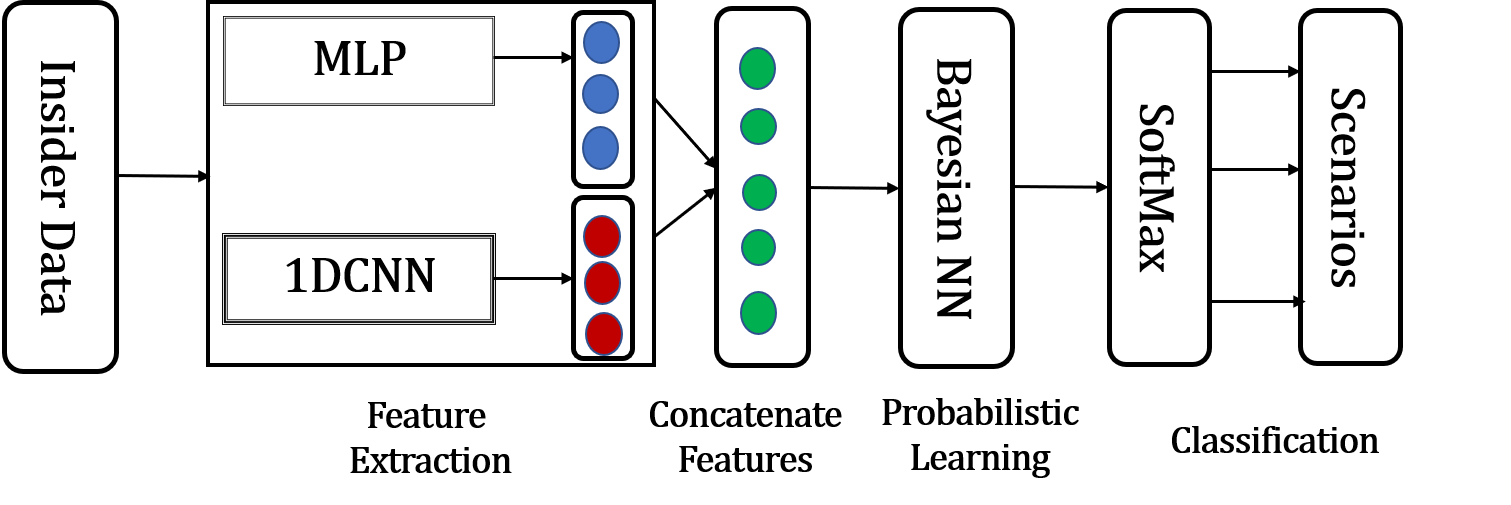}
	\caption{Hybrid Bayesian Neural Network Model}
	\label{fig:hybrid}
\end{figure}

We chose MLP and 1DCNN as the neural network algorithms for our experiment. MLP, a popular neural network, is suitable for numerical data and known for better generalization. 
In this work, we propose a hybrid BNN with \textit{\textit{(D + P)}} layers where there are \textbf{\textit{D}} deterministic layers and \textbf{P} probabilistic layers. Fig. \ref{fig:hybrid} provides the proposed hybrid learning model for anomaly detection.

The deterministic layers are used for learning task-specific representations, whereas the probabilistic layers are intended for prediction formulation and uncertainty assessment using the task-specific representation. The separation of representation learning and prediction generation is common in typical deep neural networks that use the last few dense layers for prediction generation. It is natural and straightforward to convert these typical deep neural networks to hybrid BNNs by just changing the last few dense layers to probabilistic dense layers.


Bayesian approaches are linked to stochastic learning algorithms in deep learning, and this connection is used to approximate the posterior in complex models. Dropout is a very effective stochastic regularization method for DNNs. The proposed model's MLP and 1DCNN layers conduct feature extraction, while dense layers are employed to flatten the extracted features by the kernels in all layers. The Softmax layer computes categorical probabilities for organizing the final output. Softmax is used in the multiclassification technique to transform the output of multiple neurons to the (0, 1) classification interval. We carried out experiments to demonstrate the efficacy of the proposed approach.


\subsection{Synthetic Data Evaluation}\label{sec_syn_data_eval}

Recent advances in generative modeling have highlighted the need for adequate quantitative and qualitative techniques to evaluate trainable models. Not only can reliable assessment measures aid in the evaluation of GAN models, but they also aid in discovering any inaccuracies in the data they generate. When people struggle to determine the quality of synthetic data, such as medical imaging, the requirement for acceptable measures becomes critical. It is not easy to assess a GAN model because different measures can yield different outcomes, and a high score on one evaluation criterion does not guarantee a high score on another \cite{gan_eval_metrics}.

As discussed in \cite{alaa2021faithfulganeval}, the performance of a generative model is represented as a three-dimensional point, with each dimension corresponding to different model characteristics, namely fidelity, diversity, and generalization. Fidelity refers to the quality of a model's synthetic samples, while diversity indicates how effectively these samples represent the extensive range of real samples, and generalization indicates how well a model overfits the training data.

Despite the measures like Inception Score (IS), Frechet Inception Distance (FID) FD, etc available for image data, the tabular data analysis is still an open research direction. In this work, we use multiple measures like statistical, clustering-based and manifold learning based metrics to analyze the fidelity. We use a Similarity Metric proposed by \cite{brenninkmeijer2019tabevaluator} which is an aggregation for the most prominent measures that describe the multivariate data distribution. The Silhouette coefficient \cite{rousseeuw1987silhouettes} measures cluster cohesion and separation that quantifies how well a data point fits into its allocated cluster based on two factors: how near it is to other data points in the cluster and how far it is from points in other clusters. We compare real and synthetic data using the Silhouette score, which reveals information about mode collapse. Moreover, we exploit the SPCA \cite{krzanowski1979between} to compare the data similarity. 

Diversity of data is presented using visualization of the data generated from synthetic models. We use the visual methods from t-SNE and kernel density estimation (KDE) to show the similarity of the original and synthetic data. Machine Learning Efficiency has been popularly used as a generalization metric for generative models. The generalization is tested using the synthetic data on various deep learning models for anomaly detection. Moreover, the robustness analysis also provides the efficiency of synthetic data on various other data poisoning attacks and deep learning models. The next section gives the detailed validation of the proposed approach.

\subsection{Robustness and adversarial attacks}

Adversarial robustness \cite{szegedy2013adv_robustness}, \cite{bai2021recentadv_robustness} refers to a model's performance on test data when subjected to adversarial attacks, i.e., adversarial test classification accuracy. We expect to know the model's performance in the worst-case scenario; thus, the adversary should be capable of launching attacks in white-box scenarios, in which the opponent has complete information of the model, including architectures, parameters, training data, and so on. Continuously inputting various types of adversarial samples and performing adversarial training improves the robustness of a deep network.


Machine learning algorithms accept numerical feature vectors as input. An adversarial attack is when you design input in a certain way to elicit the wrong predictions from the model. Researchers have found that adversarial examples transfer effectively between models, meaning that they can be built for a model but be effective against any other model trained on the same dataset. This is the transferability property of adversarial samples, which attackers can use when they don't have access to all the model's details.

We investigate adversarial attacks in a white box situation, in which the adversary has complete knowledge of the training process – specifically, the type of classifier employed, the training data, and any settings – but cannot influence training in any way. We use one-step attacks to examine how adversarially trained models perform. The adversarial samples are created from the distribution of test data and fed into the classifiers. The labels of the malicious samples have been reversed. Furthermore, if the learning model is used in an adversarial environment, it must be trained using adversarial approaches. Using the test split from the original data, we construct a target MLP model. The malicious instances are used to train the target classifier. The malicious instances of the test split are used to train the target classifier, ensuring that there are samples from all malicious scenarios. This target model is used to generate adversarial samples of malicious scenarios during validation. These samples are given random labels and injected into the test data with the original data. Specifically, the test data is trained by malicious samples with reversed labels. Here, we use white-box attacks like Fast Gradient Sign Method (FGSM) \cite{fgsm}, and DeepFool \cite{moosavi2016deepfool} to generate the adversarial samples and study the impact on classifiers and provide the results in Section \ref{sec_robustness}.

\section{Experiments}\label{sec_exp_setup}

In this section, we evaluate the performance of the proposed method. We provide the dataset description followed by the models being used for training and performance evaluation. We have implemented the method using Python programming language, Tensorflow and Keras. Bayesian learning is implemented using TensorFlow Probability, a library for probabilistic modeling and inference which, with Keras / TensorFlow.

\subsection{Dataset Description}
There is a scarcity of real-world data for insider threat detection. Usual practice is to use either data collected from real user data, or use synthetically generated data. Malicious insiders are predominantly employees in an organization with access rights. Data collection involves direct tracking and monitoring of the user actions and behaviors of the employees. This creates privacy and confidentiality concerns in an organization.  

This work uses the CERT insider threat datasets, which are publicly available datasets used for research, development, and testing of insider threat mitigation strategies.   We primarily choose dataset releases 4.2 (CERT r4.2) and 5.2 (CERT r5.2), which model organizations with 1000 and 2000 employees over 18 months, respectively. CERT datasets for the data collecting process (Section 4) include user activity logs in the following categories: log on/off, email, web, file and thumb drive connect, organizational structure, and user information.
A variety of models, including topic, behavior, and psychometric models, are used to provide data that is as similar to reality as possible. CERT r4.2 has three malicious insider threat scenarios (1, 2, and 3), while CERT r5.2 includes an additional scenario (4). Each malicious insider in the CERT data falls under one of four typical insider threat scenarios: data exfiltration (scenario 1), intellectual property theft (scenarios 2–4), and IT sabotage (scenarios 3–5). (scenario 3). Insider scenarios are described in detail in \cite{cmu}. Table \ref{tab:dataset} represents the summary of activity logs used for pre-processing.

%
%

\begin{table}[!hbtp]
	\centering
	\caption{Summary of the heterogeneous log files}
	\label{tab:dataset}
	\begin{tabular}{lcc}
		\hline
		\multicolumn{1}{c}{\textbf{Properties}} & \textbf{Version 4.2}  & \textbf{Version 5.2} \\ \hline
		No.of users                             & 1000                  & 2000                 \\ 
		No.of Insiders                          & 70                    & 99                   \\ 
		No.of Scenarios                         & 3                     & 4                    \\ \hline
		\multicolumn{1}{c}{\textbf{Data}}       & \multicolumn{2}{c}{\textbf{No.of Instances}} \\ \hline
		Logon.csv                               & 854859                & 1810070              \\ 
		Email.csv                               & 2629979               & 17361575             \\ 
		Http.csv                                & 1048575               & 58960449             \\ 
		Device.csv                              & 405380                & 836984               \\ 
		File.csv                                & 445581                & 887621               \\ \hline
		\multicolumn{3}{c}{\textbf{Pre-processed Data Summary}}                                \\ \hline
		Non-malicious events                    & 7154815               & 1436566              \\ 
		Malicious events                        & 7323                  & 10329                \\ 
		Total instances                         & 330452                & 1048575              \\ 
		Malicious                               & 966                   & 1132                 \\ 
		Non-Malicious                           & 329487                & 1047443              \\ \hline
	\end{tabular}
\end{table}


\subsection{Performance Evaluation}
In this section, we discuss the metrics and measures used for the performance evaluation of GAN and the anomaly detection used for insider threat detection. As discussed in Section \ref{sec_syn_data_eval}, we provide the results for fidelity, diversity and generalization of the GAN data. The generalization of the GAN data is tested using the machine learning efficacy - how the GAN data performs in machine learning models. Deep learning models are used for the anomaly detection. Hence, generalization is validated based on the performance of the GAN data on these models. The classification models are validated using commonly used performance metrics for imbalanced data, such as precision (P), recall (R), and f-score (F). 

Recall, also referred as Detection Rate (DR), is a measure of a classifier’s completeness. The higher the recall, the more cases the classifier covers. If the DR is larger, the classification performance of the model is better. It is best suggested to have high precision and recall, and these metrics alone do not always accurately reflect the actual performance of various models. The majority class is typically regarded as the negative class, and it is due to the assumption that interesting samples are uncommon and thus considered positive. 

Even though previous works achieved high precision and recall, no work has performed a detailed analysis of learning reports. We validated the results against the number of false positives and false negatives and analyzed them based on precision and recall. Given the importance of the confusion matrix, also known as an error matrix, we have included two additional metrics for the experimental evaluation: Cohen's Kappa (Kappa) and Mathews Correlation Coefficient (MCC).

Cohen's Kappa, also known as the Kappa score, measures the degree of agreement between the true and predicted values. Cohen's Kappa is always less than or equal to 1, and a value of 0 or less indicates that the classifier is ineffective. Kappa score between 0.81–1 indicates nearly perfect agreement; the score reaches its maximum for balanced data. The Cohen’s kappa is the classification accuracy normalized by the imbalance of the classes in the data.

Precision, recall, and F-score are asymmetric; their values change when the classes are switched. The positive class is the class of interest for precision and recall. The confusion matrix's True-Positive (TP), False-Positive (FP), and False-Negative (FN) values are used to compute them. These metrics do not use True-Negative (TN). Changes in TN never reflect changes in precision and recall. As a result, we consider MCC, a symmetric metric with a value between -1 and +1 inclusive. MCC considers all four values in the confusion matrix; no class is more important than the other, so switching between the negative and positive classes yields the same value. A value close to 1 indicates that both classes are accurately predicted. The experimental results are detailed in the Section \ref{sec_results}.

\subsection{Experimental Results}\label{sec_results}
In this section, we explain the performance of multi-class anomaly detection approach using deep learning algorithms for insider detection. Comprehensive set of experiments were conducted on original data as well as applying various data augmentation methods as follows : (i) on original skewed data with no data augmentation (ii) on CGAN augmented training data (iii) on CWGAN-GP augmented training data (iv) on ACGAN augmented training data and (v) SPCAGAN generated synthetic data in the training set. We used the algorithms MLP, 1DCNN and Bayesian Neural Network individually, ensemble of MLP and 1DCNN and a hybrid model using MLP and 1DCNN ensemble on BNN. These experiments helped in analyzing the behavior of data in various model setting. 

ML/DL methods were used for building anomaly detection models on all the above mentioned data settings. We use text highlighted in bold to depict the interesting results in the tables with experimental results. The tables use the abbreviations Precision (P), Recall (R), F-score (F), Cohen's Kappa (K) and Mathews Correlation Coefficient (MCC). Table \ref{tab:multiclass} gives the performance of anomaly detection using multi-class classification.

\begin{table*}[!hbtp]
	\centering
	\caption{Performance comparison of various deep learning models for insider threat detection}
	\label{tab:multiclass}
	\begin{tabular}{@{}lllllllllll@{}}
		\toprule
		\multicolumn{1}{c}{\multirow{2}{*}{\textbf{Method}}} & \multicolumn{5}{c}{\textbf{CERT v4.2}}                                                                                                                                & \multicolumn{5}{c}{\textbf{CERT v5.2}}                                                                                                                                \\ \cmidrule(l){2-11} 
		\multicolumn{1}{c}{}                                 & \multicolumn{1}{c}{\textbf{P}} & \multicolumn{1}{c}{\textbf{R}} & \multicolumn{1}{c}{\textbf{F}} & \multicolumn{1}{c}{\textbf{K}} & \multicolumn{1}{c}{\textbf{MCC}} & \multicolumn{1}{c}{\textbf{P}} & \multicolumn{1}{c}{\textbf{R}} & \multicolumn{1}{c}{\textbf{F}} & \multicolumn{1}{c}{\textbf{K}} & \multicolumn{1}{c}{\textbf{MCC}} \\ \midrule
		\textbf{Real +   MLP}                                & 0.2750                         & 0.7820                         & 0.2830                         & 0.0550                         & 0.1680                           & 0.3085                         & 0.4133                         & 0.3366                         & 0.1108                         & 0.1145                           \\ 
		\textbf{CGAN +   MLP}                                & 0.5167                         & 0.9087                         & 0.5306                         & 0.0639                         & 0.1653                           & 0.5392                         & 0.7612                         & 0.5531                         & 0.1605                         & 0.2920                           \\ 
		\textbf{CWGAN-GP   + MLP}                            & 0.6495                         & 0.9528                         & 0.7237                         & 0.4483                         & 0.5204                           & 0.5245                         & 0.5794                         & 0.4883                         & 0.7013                         & 0.7034                           \\ 
		\textbf{ACGAN+MLP}                                   & 0.7582                         & 0.6681                         & 0.6777                         & 0.7790                         & 0.7863                           & 0.6360                         & 0.7793                         & 0.6530                         & 0.5837                         & 0.6050                           \\ 
		\textbf{SPCAGAN   + MLP}                             & \textbf{0.7897}                & \textbf{0.7175}                & \textbf{0.7467}                & \textbf{0.8658}                & \textbf{0.8701}                  & \textbf{0.7909}                & \textbf{0.7496}                & \textbf{0.7583}                & \textbf{0.8727}                & \textbf{0.8729}                  \\ \midrule
		\textbf{Real +   1DCNN}                              & 0.2750                         & 0.7880                         & 0.2830                         & 0.0550                         & 0.1680                           & 0.4420                         & 0.7427                         & 0.1274                         & 0.0031                         & 0.0380                           \\ 
		\textbf{CGAN +   1DCNN}                              & 0.5787                         & 0.9902                         & 0.6324                         & 0.2686                         & 0.3929                           & 0.4993                         & 0.4559                         & 0.4755                         & 0.1344                         & 0.2601                           \\ 
		\textbf{CWGAN-GP   + 1DCNN}                          & 0.5053                         & 0.9306                         & 0.5051                         & 0.0205                         & 0.0954                           & 0.6667                         & 0.7500                         & 0.7000                         & 0.4000                         & 0.4082                           \\ 
		\textbf{ACGAN+1DCNN}                                 & 0.6437                         & 0.6562                         & 0.6489                         & 0.7487                         & 0.7497                           & 0.6360                         & 0.7793                         & 0.6530                         & 0.5837                         & 0.6050                           \\ 
		\textbf{SPCAGAN   + 1DCNN}                           & \textbf{0.7691}                & \textbf{0.7322}                & \textbf{0.7293}                & \textbf{0.8544}                & \textbf{0.8571}                  & \textbf{0.6703}                & \textbf{0.5128}                & \textbf{0.5748}                & \textbf{0.4359}                & \textbf{0.4398}                  \\ \midrule
		\textbf{Real +   Ensemble}                           & 0.2980                         & 0.9852                         & 0.3313                         & 0.0949                         & 0.2228                           & 0.3397                         & 0.2699                         & 0.2931                         & 0.4343                         & 0.4836                           \\ 
		\textbf{CGAN +   Ensemble}                           & 0.7541                         & 0.7480                         & 0.7393                         & 0.9204                         & 0.9206                           & 0.6437                         & 0.6562                         & 0.6489                         & 0.7487                         & 0.7497                           \\ 
		\textbf{CWGAN-GP   + Ensemble}                       & 0.8333                         & 0.7500                         & 0.7857                         & 0.5714                         & 0.5773                           & 0.7072                         & 0.7114                         & 0.7047                         & 0.8400                         & 0.8403                           \\ 
		\textbf{ACGAN +   Ensemble}                          & 0.6409                         & 0.6429                         & 0.6417                         & 0.8515                         & 0.8517                           & 0.6724                         & 0.4317                         & 0.4618                         & 0.4875                         & 0.5491                           \\ 
		\textbf{SPCAGAN+Ensemble}                            & \textbf{0.7151}                & \textbf{0.6579}                & \textbf{0.6841}                & \textbf{0.8769}                & \textbf{0.8815}                  & \textbf{0.7926}                & \textbf{0.5637}                & \textbf{0.5393}                & \textbf{0.6771}                & \textbf{0.6875}                  \\ \midrule
		\textbf{Real +   BNN}                                & 0.3533                         & 0.3736                         & 0.3618                         & 0.3617                         & 0.3636                           & 0.3136                         & 0.5178                         & 0.3154                         & 0.0446                         & 0.1125                           \\ 
		\textbf{CGAN +   BNN}                                & 0.4899                         & 0.5116                         & 0.4851                         & 0.7995                         & 0.8008                           & 0.2262                         & 0.5648                         & 0.2366                         & 0.0193                         & 0.0854                           \\ 
		\textbf{CWGAN-GP   + BNN}                            & 0.6724                         & 0.4317                         & 0.4618                         & 0.4875                         & 0.5491                           & 0.7132                         & 0.6550                         & 0.6816                         & 0.8679                         & 0.8728                           \\ 
		\textbf{ACGAN +   BNN}                               & 0.6103                         & 0.4666                         & 0.4541                         & 0.4498                         & 0.4691                           & 0.3136                         & 0.5178                         & 0.3154                         & 0.0446                         & 0.1125                           \\ 
		\textbf{SPCAGAN+BNN}                                 & \textbf{0.6720}                & \textbf{0.7494}                & \textbf{0.6277}                & \textbf{0.3282}                & \textbf{0.4357}                  & \textbf{0.5998}                & \textbf{0.3166}                & \textbf{0.3765}                & \textbf{0.4927}                & \textbf{0.5534}                  \\ \midrule
		\textbf{Real   +Hybrid}                              & 0.4702                         & 0.4448                         & 0.4567                         & 0.8175                         & 0.8230                           & 0.5368                         & 0.3660                         & 0.4208                         & 0.4777                         & 0.5241                           \\ 
		\textbf{CGAN   +Hybrid}                              & 0.7691                         & 0.7322                         & 0.7293                         & 0.8544                         & 0.8571                           & 0.5208                         & 0.3107                         & 0.3489                         & 0.4061                         & 0.4713                           \\ 
		\textbf{CWGAN-GP   + Hybrid}                         & 0.7132                         & 0.6550                         & 0.6816                         & 0.8679                         & 0.8728                           & 0.7799                         & 0.3433                         & 0.4244                         & 0.1784                         & 0.2277                           \\ 
		\textbf{ACGAN +   Hybrid}                            & 0.6906                         & 0.6154                         & 0.6491                         & 0.8364                         & 0.8430                           & 0.6724                         & 0.4317                         & 0.4618                         & 0.4875                         & 0.5491                           \\ 
		\textbf{SPCAGAN+Hybrid}                              & \textbf{0.8956}                & \textbf{0.6953}                & \textbf{0.7385}                & \textbf{0.8498}                & \textbf{0.8509}                  & \textbf{0.8754}                & \textbf{0.9762}                & \textbf{0.9198}                & \textbf{0.8397}                & \textbf{0.8457}                  \\ \bottomrule
	\end{tabular}
\end{table*}


The deep learning algorithms need more data to learn from the deep representative features. Hence, MLP and 1DCNN show remarkable improvement in the metrics. The performance of SPCAGAN augmented data for the anomaly detection is evident from the striking improvement in the metrics as shown in Table \ref{tab:multiclass}. The results offer compelling evidence for supporting the efficiency of our proposed method.

\subsection{Robustness Analysis}\label{sec_robustness}
In this section, we analyze the behavior of trained models in the presence of adversarial samples. The robustness of adversarial training is evaluated by creating samples using adversarial data generation methods like FGSM and DeepFool. The model trained using GAN augmented training data should be efficient when tested with FGSM and DeepFool generated adversarial samples. The already trained model is tested with these examples. FGSM \cite{fgsm} combines a white-box approach with a misclassification goal, and it tricks a neural network model into making wrong predictions.

We perform the experiments for the robustness to the white-box attacks in the following setting. Table \ref{tab:classification} in Section \ref{sec_results} provided the performance analysis for the various linear and non-linear classification models in the adversarial trained data. These models are validated against the adversarial examples generated using FGSM and DeepFool attacks and the results are given in Table \ref{tab:robustness}. 

\begin{table*}[!hbtp]
	\centering
	\caption{Performance analysis of FGSM and DeepFool attacks on the deep learning models trained using SPCAGAN augmented insider threat data}
	\label{tab:robustness}
	\begin{tabular}{@{}llllllllllll@{}}
		\toprule
		\multirow{2}{*}{\textbf{Attack}} & \multicolumn{1}{c}{\multirow{2}{*}{\textbf{DL Model}}} & \multicolumn{5}{c}{\textbf{CERT   v4.2}}                                                                                                                           & \multicolumn{5}{c}{\textbf{CERT   v5.2}}                                                                                                                           \\ \cmidrule(l){3-12} 
		& \multicolumn{1}{c}{}                                   & \multicolumn{1}{c}{\textbf{P}} & \multicolumn{1}{c}{\textbf{R}} & \multicolumn{1}{c}{\textbf{F}} & \multicolumn{1}{c}{\textbf{K}} & \multicolumn{1}{c}{\textbf{M}} & \multicolumn{1}{c}{\textbf{P}} & \multicolumn{1}{c}{\textbf{R}} & \multicolumn{1}{c}{\textbf{F}} & \multicolumn{1}{c}{\textbf{K}} & \multicolumn{1}{c}{\textbf{M}} \\ \midrule
		\multirow{4}{*}{\textbf{FGSM}}   & \textbf{MLP}                                 & 0.4084                         & 0.3076                         & 0.3187                         & 0.1506                         & 0.1530                         & 0.2631                         & 0.8988                         & 0.2293                         & 0.0129                         & 0.0796                         \\ 
		& \textbf{1DCNN}                               & 0.5749                         & 0.3680                         & 0.4313                         & 0.4970                         & 0.5659                         & 0.5468                         & 0.5558                         & 0.4709                         & 0.6414                         & 0.6423                         \\ 
		& \textbf{Ensemble}                            & 0.6103                         & 0.4666                         & 0.4541                         & 0.4498                         & 0.4691                         & 0.6807                         & 0.4638                         & 0.4967                         & 0.2701                         & 0.2750                         \\ 
		& \textbf{Hybrid}                          & 0.8162                         & 0.5790                         & 0.6670                         & 0.6362                         & 0.6725                         & 0.7781                         & 0.5931                         & 0.5227                         & 0.7135                         & 0.7152                         \\ \midrule
		\multirow{4}{*}{\textbf{DF}}     & \textbf{MLP}                                 & 0.2725                         & 0.6507                         & 0.2743                         & 0.0349                         & 0.0951                         & 0.4966                         & 0.5359                         & 0.4640                         & 0.7313                         & 0.7316                         \\ 
		& \textbf{1DCNN}                               & 0.6159                         & 0.5040                         & 0.5249                         & 0.5973                         & 0.6404                         & 0.3085                         & 0.4133                         & 0.3386                         & 0.1108                         & 0.1145                         \\ 
		& \textbf{Ensemble}                            & 0.6511                         & 0.4833                         & 0.5287                         & 0.5122                         & 0.5240                         & 0.5745                         & 0.3576                         & 0.3905                         & 0.1639                         & 0.1688                         \\  
		& \textbf{Hybrid}                          & 0.7822                         & 0.5450                         & 0.6330                         & 0.6022                         & 0.6385                         & 0.8293                         & 0.6443                         & 0.5739                         & 0.7647                         & 0.7664                         \\ \bottomrule
	\end{tabular}
\end{table*}

We observe that models built using original and SPCAGAN augmented data are more resistant to FGSM one-step attack. We have observed this from the experiments with varying numbers of generated samples. The experiments show a significant drop in the performance metrics with DF attacks. The model with SPCAGAN augmentation applied on hybrid BNN yields better robustness on both datasets in deep learning models. The following section shows the effectiveness of SPCAGAN-based synthetic data compared to the other data generation methods.

\subsection{Comparison with other data generation methods}
There are various data generation methods available. In this section, we discuss some of the widely used data generation approaches like Random Over Sampling (ROS) \cite{batista2004studyROS}, Synthetic Minority Oversampling Technique (SMOTE) \cite{chawla2002smote}, adding random noise to the data points \cite{hoyer2008randomnoise}, using Gaussian Mixture Models (GMM) \cite{duda1973patternGMM}. We compare the performance of these techniques against the proposed method. To validate the proposed contributions, we perform the experimental analysis of the proposed approach and the comparison against other widely used augmentation techniques.

Oversampling refers to creating data samples from the minority classes to balance the data. Random Oversampling (ROS) and Synthetic Minority Oversampling Technique (SMOTE) are two popularly used oversampling approaches. Random Oversampling (ROS) is a data augmentation approach that selects random samples from the minority class with replacement and augments the training data with various copies of these instances so that a single instance can be selected multiple times. Random oversampling may enhance the likelihood of overfitting because it simply replicates minority class samples to increase sample count. A symbolic classifier, for example, may generate rules that appear to be accurate but only consider samples generated from one class. 

SMOTE (Synthetic Minority Oversampling Technique) works by selecting a point at random from the minority class and computing its k-nearest neighbors. The synthetic points are added between the chosen point and its neighbors. The major limitation of using oversampling methods is that it increase the chances of overfitting as they replicate minority class events. The issue of oversampling is evident from the performance metrics given in the experimental results. 

Gaussian mixture models (GMMs) are based on the notion that a finite number of Gaussian distributions exist, each of which denotes a cluster. As a result, a Gaussian Mixture Model tends to group data points from a single distribution. Gaussian Mixture Models are probabilistic models that use soft clustering to distribute points into multiple clusters. This method includes modeling data with a multivariate Gaussian distribution; the parameters are then computed using Maximum Likelihood Estimation (MLE), which is based on the mean and covariance matrix of the training data.


\begin{table*}[!hbtp]
	\centering
	\caption{Synthetic Data generation using methods other than GAN}
	\label{tab:classification}
	\begin{tabular}{lllllllllll}
		\hline
		\multicolumn{1}{c}{\multirow{3}{*}{\textbf{Method}}} & \multicolumn{5}{c}{\multirow{2}{*}{\textbf{CERT v4.2}}}                                                                                                            & \multicolumn{5}{c}{\multirow{2}{*}{\textbf{CERT v5.2}}}                                                                                                            \\
		\multicolumn{1}{c}{}                                 & \multicolumn{5}{c}{}           & \multicolumn{5}{c}{}           \\ \cline{2-11} 
		\multicolumn{1}{c}{}                                 & \multicolumn{1}{c}{\textbf{P}} & \multicolumn{1}{c}{\textbf{R}} & \multicolumn{1}{c}{\textbf{F}} & \multicolumn{1}{c}{\textbf{K}} & \multicolumn{1}{c}{\textbf{M}} & \multicolumn{1}{c}{\textbf{P}} & \multicolumn{1}{c}{\textbf{R}} & \multicolumn{1}{c}{\textbf{F}} & \multicolumn{1}{c}{\textbf{K}} & \multicolumn{1}{c}{\textbf{M}} \\ \hline
		\textbf{ROS + MLP}                                   & 0.4380                         & 0.6580                         & 0.4790                         & 0.3470                         & 0.4310                         & 0.4738                         & 0.6584                         & 0.4787                         & 0.3470                         & 0.4308                         \\ 
		\textbf{ROS + 1DCNN}                                 & 0.3680                         & 0.8660                         & 0.4190                         & 0.1560                         & 0.2900                         & 0.3136                         & 0.5178                         & 0.3154                         & 0.0446                         & 0.1125                         \\ 
		\textbf{ROS + BNN}                                   & 0.4060                         & 0.7900                         & 0.4570                         & 0.1620                         & 0.2960                         & 0.4821                         & 0.6449                         & 0.4824                         & 0.4646                         & 0.5120                         \\ 
		\textbf{ROS + Hybrid}                                & 0.5357                         & 0.6249                         & 0.5555                         & 0.1110                         & 0.1335                         & 0.3683                         & 0.8514                         & 0.4208                         & 0.3158                         & 0.4284                         \\ \hline
		\textbf{Random Noise + MLP}                           & 0.3085                         & 0.4133                         & 0.3386                         & 0.1108                         & 0.1145                         & 0.5050                         & 0.4531                         & 0.4458                         & 0.6182                         & 0.6292                         \\ 
		\textbf{Random Noise + 1DCNN}                         & 0.4907                         & 0.6409                         & 0.4579                         & 0.2979                         & 0.3797                         & 0.2682                         & 0.6360                         & 0.2659                         & 0.0477                         & 0.1308                         \\ 
		\textbf{Random Noise + BNN}                           & 0.3533                         & 0.3736                         & 0.3618                         & 0.3617                         & 0.3636                         & 0.5022                         & 0.9434                         & 0.4905                         & 0.0083                         & 0.0626                         \\ 
		\textbf{Random Noise + Hybrid}                        & 0.5167                         & 0.9087                         & 0.5306                         & 0.0639                         & 0.1653                         & 0.5100                         & 0.5087                         & 0.4968                         & 0.8262                         & 0.8318                         \\ \hline
		\textbf{SMOTE + MLP}                                 & 0.5960                         & 0.7350                         & 0.6480                         & 0.7400                         & 0.7460                         & 0.4821                         & 0.6449                         & 0.4824                         & 0.4646                         & 0.5120                         \\ 
		\textbf{SMOTE + 1DCNN}                               & 0.3960                         & 0.7710                         & 0.4310                         & 0.1200                         & 0.2500                         & \multicolumn{1}{c}{0.2210}     & \multicolumn{1}{c}{0.5484}     & \multicolumn{1}{c}{0.2349}     & \multicolumn{1}{c}{0.0393}     & \multicolumn{1}{c}{0.1070}     \\ 
		\textbf{SMOTE + BNN}                                 & 0.4520                         & 0.7740                         & 0.4920                         & 0.1560                         & 0.2890                         & 0.2667                         & 0.9316                         & 0.2634                         & 0.0364                         & 0.1357                         \\ 
		\textbf{SMOTE + Hybrid}                              & 0.5270                         & 0.7780                         & 0.5240                         & 0.2330                         & 0.3600                         & 0.4993                         & 0.4559                         & 0.4755                         & 0.1344                         & 0.2601                         \\ \hline
		\textbf{GMM + MLP}                                   & 0.6862                         & 0.7170                         & 0.6765                         & 0.7333                         & 0.7389                         & 0.2980                         & 0.9852                         & 0.3313                         & 0.0949                         & 0.2228                         \\ 
		\textbf{GMM + 1DCNN}                                 & 0.5734                         & 0.5260                         & 0.4964                         & 0.6262                         & 0.6523                         & 0.2748                         & 0.7891                         & 0.2833                         & 0.0547                         & 0.1676                         \\ 
		\textbf{GMM + BNN}                                   & 0.4052                         & 0.3505                         & 0.3087                         & 0.1679                         & 0.1743                         & 0.3788                         & 0.9175                         & 0.4413                         & 0.3732                         & 0.4766                         \\ 
		\textbf{GMM + Hybrid}                                & 0.6250                         & 0.7500                         & 0.6666                         & 0.3333                         & 0.3535                         & 0.2500                         & 0.6250                         & 0.2667                         & 0.3333                         & 0.3535                         \\ \hline
	\end{tabular}
\end{table*}

Table \ref{tab:classification} provides the performance analysis of using various other methods for data augmentation. Overall better results are highlighted. Compared to Table \ref{tab:multiclass}, the results in Table \ref{tab:classification} show that no method provided agreeable performance for insider threat detection. GMM could perform better when compared to other methods on v4.2. But for v5.2, where the class imbalance is extreme, GMMs failed to generate quality data and hence poor performance in learning models.In the following section, we provide the authenticity evaluation of synthetic data generated using the proposed SPCAGAN.

\subsection{Authenticity of Synthetic Data}
This section deals with the authenticity of the GAN generated synthetic data. As discussed in Section \ref{sec_syn_data_eval}, we validated the synthetic data from three criteria : Fidelity, Diversity and the Generalization. We consider three metrics - SPCA, Similarity Score (SS) and Silhouette Coefficient (SC) for measuring the fidelity. Diversity is validated using visual methods like PCA and t-SNE. For generalization, we used the machine learning efficiency and is discussed in Section \ref{sec_results}. 

\begin{figure}[tbph!]
	\centering
	\includegraphics[scale=0.35]{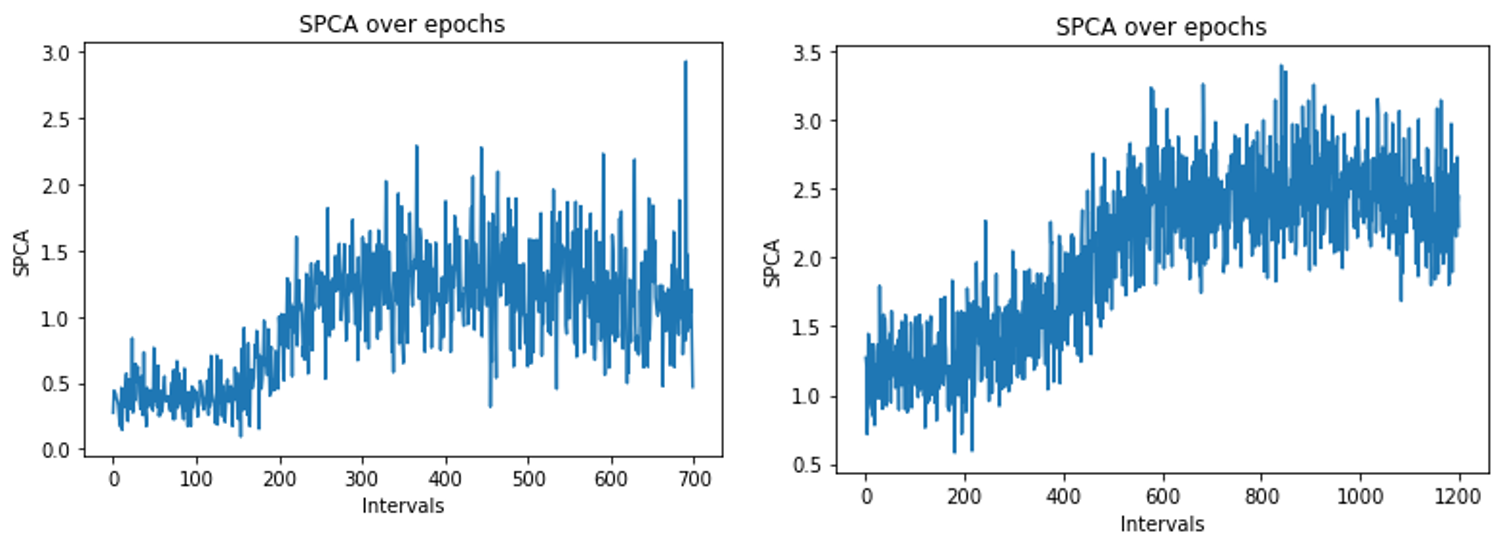}
	\caption{SPCA plot during GAN training CERT v4.2 and v5.2 with SPCAGAN}
	\label{fig:spca}
\end{figure}

The SPCAGAN training is monitored in frequent intervals to study how the SPCA values progress. PCA is performed on both the datasets to find the principal components needed for maximum information utilization. In general, the number of principal components (PC) created will be the minimum value found of either (n-1) where n is the number of data points in a dataset or (p) where p is the number of variables in your data. We used the elbow method, a common heuristic in mathematical optimization, to get the number of principal components to be used in the SPCA calculation. CERT v4.2 requires three PCs whereas CERT v5.2 requires four. SPCA intuitively measures the similarity of two matrices by computing the squared cosine values of all the combinations of the first k principal components of two matrices. Fig. \ref{fig:spca} shows the SPCA values changing during the GAN training for CERT v4.2 and CERT v5.2.

Table \ref{tab:fidelity} compares the fidelity measures for SPCA, similarity metric and the Silhouette score for the similarity between real and synthetic data from various data generation approaches.

From the table, it is evident that the SPCAGAN provided the best similarity between the original and the synthetic data. We considered all the three metrics for comparison. The compared methods include popular Random oversampling, SMOTE, Random noise addition, and other popular GAN methods like CGAN, CWGAN-GP and ACGAN.

\begin{table}[!hbtp]
	\centering
	\caption{Fidelity measures for synthetic data generation methods}
	\begin{tabular}{@{}lcccccc@{}}
		\toprule
		\multirow{2}{*}{\textbf{Method}} & \multicolumn{3}{c}{\textbf{CERT v4.2}}           & \multicolumn{3}{c}{\textbf{CERT v5.2}}          \\ \cmidrule(l){2-7} 
		& \textbf{SPCA} & \textbf{SS}    & \textbf{SC}    & \textbf{SPCA} & \textbf{SS}    & \textbf{SC}    \\ \midrule
		\textbf{ROS}                     & 0.985         & 0.281          & 0.315          & 0.874         & 0.250          & 0.386          \\ 
		\textbf{SMOTE}                   & 1.364         & 0.459          & 0.399          & 1.035         & 0.378          & 0.298          \\ 
		\textbf{GMM}                     & 1.726         & 0.483          & 0.582          & 1.617         & 0.543          & 0.313          \\ 
		\textbf{RN}                      & 1.873         & 0.461          & 0.433          & 1.806         & 0.461          & 0.311          \\ 
		\textbf{CGAN}                    & 2.235         & 0.796          & 0.539          & 2.187         & 0.826          & 0.541          \\ 
		\textbf{CWGAN-GP}                & 2.421          & 0.905          & 0.624          & 2.263         & 0.467          & 0.674          \\ 
		\textbf{ACGAN}                   & 2.438          & 0.891          & 0.588          & 2.804         & 0.685          & 0.621          \\
		\textbf{SPCAGAN}                 & \textbf{2.973} & \textbf{0.978} & \textbf{0.642} & \textbf{3.868} & \textbf{0.917} & \textbf{0.668} \\ \bottomrule
	\end{tabular}
	\label{tab:fidelity}
\end{table}

Diversity of the data refers to the ability of the GAN model to generate synthetic data without mode collapse. Metrics for evaluating the similarity of synthetic data to original data distributions for numerical data types are not well developed. In the case of images, metrics such as Inception Score are used to compare synthetic and real images. Due to the lack of a proven metric for comparing the numerical data similarity between the original and synthetic data distributions, we utilize a visual approach based on kernel density estimates (KDE), Principal Component Analysis (PCA), and t-distributed Stochastic Neighbor Embedding (t-SNE) on the original and synthetic data.

\begin{figure}[!htbp]
	\centering
	\includegraphics[scale=0.3]{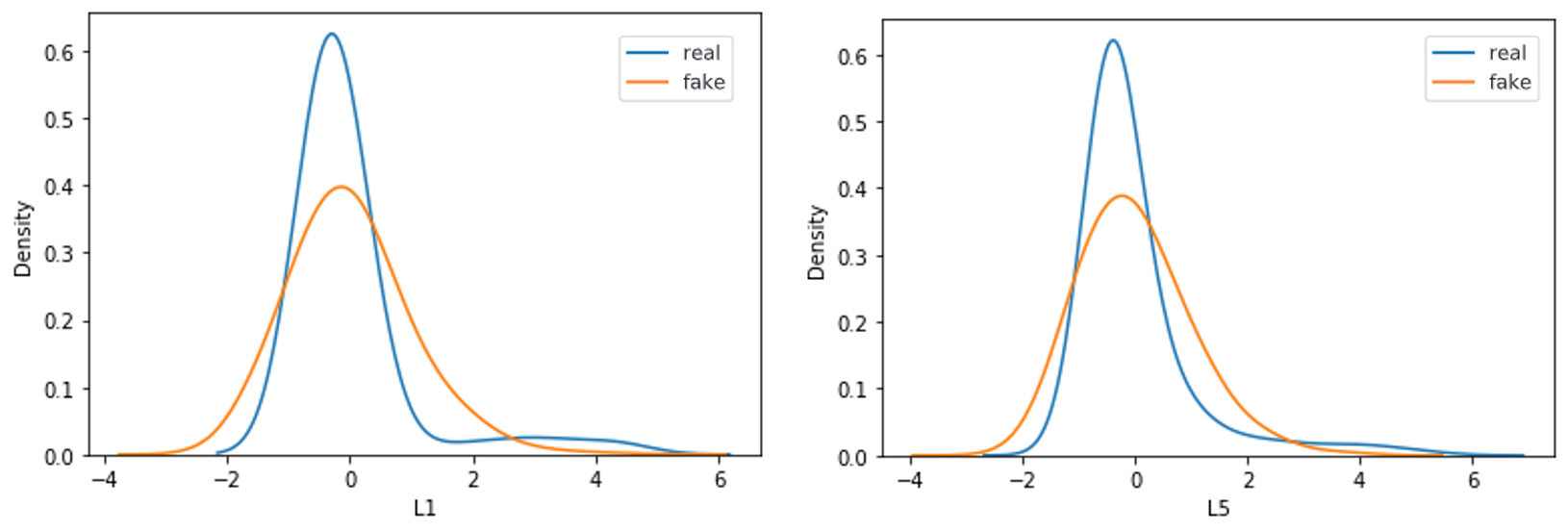}
	\caption{Kernel density estimation of original and synthetic data}
	\label{fig:kde}
\end{figure}

To demonstrate the comparability of data distributions, we employed kernel density estimation, a non-parametric approach, on both the original and synthetic data. Due to space constraints, we use the visualization to illustrate two features. Fig. \ref{fig:kde} depicts the kernel density estimation plot on two features L1 and L5.

PCA is a technique that converts n-dimensions of data into k-dimensions while maintaining as much information from the original dataset. PCA differs from t-SNE where the later one preserves the local neighbors of the data points. t-SNE can be considered as a manifold learning where the geometric data properties are used. t-SNE helps to increase the interpretability of data in the lower dimensions.

\begin{figure}[tbph!]
	\centering
	\includegraphics[scale=0.27]{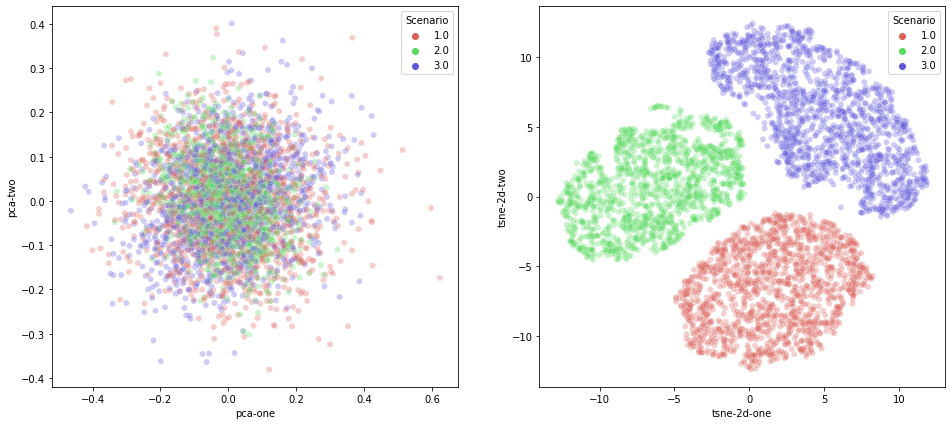}
	\caption{PCA and t-SNE based visualization for CERT v4.2}
	\label{fig:pcatsnefakev4}
\end{figure}

\begin{figure}[tbph!]
	\centering
	\includegraphics[scale=0.27]{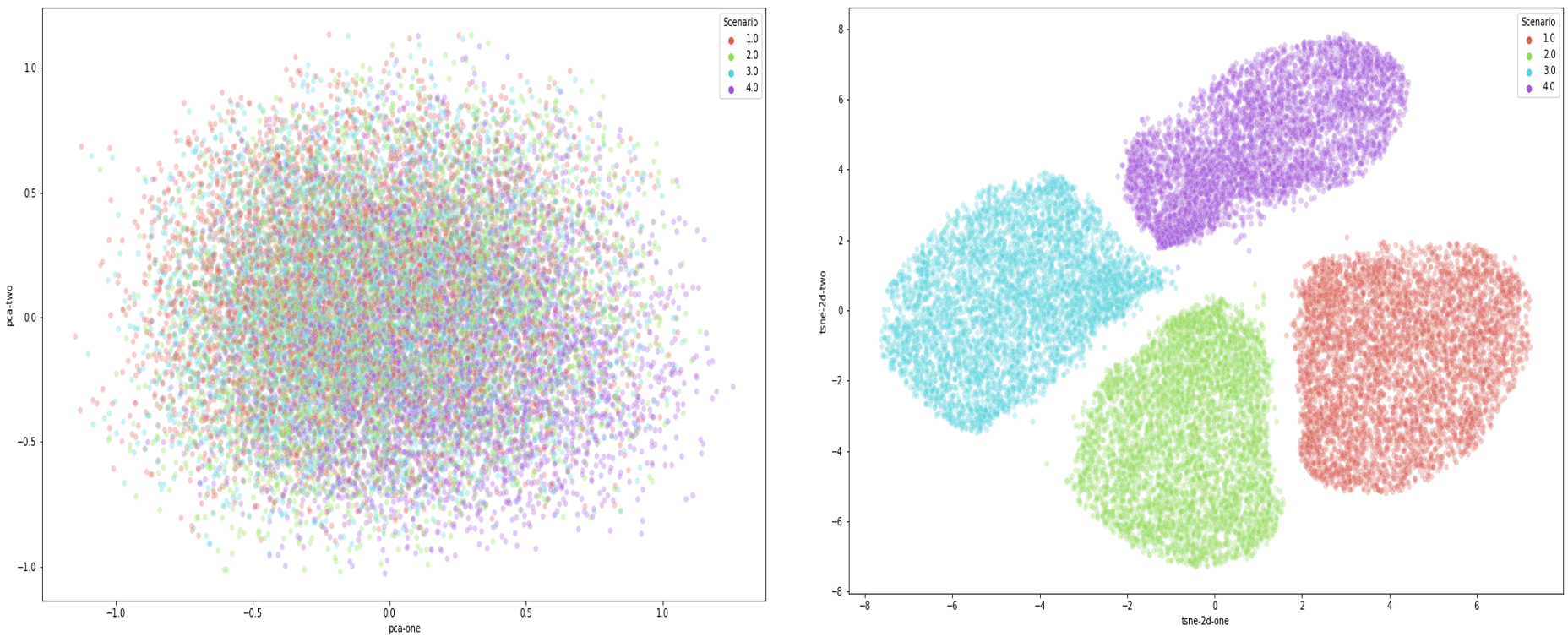}
	\caption{PCA and t-SNE based visualization for CERT v5.2}
	\label{fig:pcatsnefakev5}
\end{figure}

Figures \ref{fig:pcatsnefakev4} and \ref{fig:pcatsnefakev5} show the PCA and t-SNE visualization of the data on 2D for CERT v4.2 and CERT v5.2. PCA shows the dimension reduced data. Since it is a 2D visualization, the separate clusters are not visible; hence the overlap. In the t-SNE based visualization, the three clusters are the minority insider activity samples generated using the SPCAGAN model. CERT v4.2 displays three malicious scenarios in Fig.\ref{fig:pcatsnefakev4} whereas CERT v5.2 shows four scenarios in Fig.\ref{fig:pcatsnefakev5}. The larger data spread indicates that there is no obvious mode-collapse.

Even though the literature shows numerous solution approaches being used in insider threat attacks, comparing these methods is challenging due to differences in many aspects like the data and the feature space used for training and validation and the learning methods applied for threat analysis. We tried to incorporate different studies where insider detection was studied previously in Section \ref{sec_rel_work_insider}. Many papers used the popular ensemble XGBoost for the insider threat analysis. In this work, the main aim was to use deep learning based models. The results from the proposed approach shows significant improvement than other works. Overall, the proposed SPCAGAN based adversarial training and hybrid BNN for anomaly detection yielded expected results in the experiments. In the next section, we summarize the work by discussing the conclusion and scope for future enhancements. 


. 

\section{Conclusion}\label{sec_conclusion}
With the aim to improve the detection of minority classes in insider detection, and to reduce the effect of imbalance of class distribution in data, this article proposes a novel data augmentation based training method for insider detection SPCAGAN. SPCAGAN generates new samples of specified classes such that imbalance of certain classes in training dataset is reduced. It also ensures the diversity of samples in the training data. The research is performed as a complete workflow starting from feature space creation and feature set generation, pre-processing of data followed by labeling the samples. Next stage was SPCAGAN network design, training and data generation. Final step involved the validation of the method. The detection performance of insider threats was validated using the benchmark CMU CERT dataset, and the expected results were obtained. The results of the experiments reveal that the proposed SPCAGAN has a better detection effect on the minority classes in the imbalanced dataset. 

We have identified directions for future work. Though BNNs combine the powerful features of probabilistic modeling and neural networks, they also suffer from some of their weaknesses. Due to the sampling or variational inference phases, BNNs are currently computationally expensive; hence additional refinement and optimization are required to lower the time cost.



\end{document}